\documentclass[useAMS,usenatbib]{mn2e}

\usepackage[dvips]{graphicx}
\usepackage{setspace}
\usepackage{multirow}
\usepackage{color}
\usepackage {amssymb}

\newcommand{\apj}{ApJ}
\newcommand{\apjl}{ApJL}
\newcommand{\mnras}{MNRAS}

\newcommand{\apjs}{ApJS}
\newcommand{\nat}{Nature}
\newcommand{\araa}{ARA\&A}
\newcommand{\aap}{A\&A}
\newcommand{\aaps}{A\&AS}
\newcommand{\pasp}{PASP}

\newcommand{\ls}[1]{ %
  \dimen0=\fontdimen6\the\font                                                  
  \lineskip=#1\dimen0                                                           
  \advance\lineskip.5\fontdimen5\the\font                                       
  \advance\lineskip-\dimen0                                                     
  \lineskiplimit=.9\lineskip                                                    
  \baselineskip=\lineskip                                                       
  \advance\baselineskip\dimen0                                                  
  \normallineskip\lineskip                                                      
  \normallineskiplimit\lineskiplimit                                            
  \normalbaselineskip\baselineskip                                              
  \ignorespaces                                                                 
}%

\title[Maximum jet efficiency for powerful radio
  sources]{Evidence for a maximum jet efficiency for the
    most powerful radio galaxies} \author[C. A. C. Fernandes et
  al. ]{C. A. C. Fernandes$^{1}$\thanks{E-mail: cacf@astro.ox.ac.uk},
  M. J. Jarvis$^{2}$, S. Rawlings$^{1}$,
  A. Mart\'{i}nez-Sansigre$^{1,3}$,\newauthor E. Hatziminaoglou$^{4}$,
  M. Lacy$^{5}$, M. J. Page$^{6}$, J. A. Stevens$^{2}$ and
  E. Vardoulaki$^{1}$ \\$^{1}$University of Oxford, Subdepartment of
  Astrophysics, Denys Wilkinson Building, Keble Road, Oxford OX1 2DL,
  UK\\$^{2}$Centre for Astrophysics Research, STRI, University of
  Hertfordshire, Hatfield, AL10 9AB, UK\\$^{3}$Institute of Cosmology
  and Gravitation, University of Portsmouth, Dennis Sciama Building,
  Burnaby Road, Portsmouth, PO1 3FX, UK\\$^{4}$European Southern
  Observatory, Karl-Schwarzschild-Str. 2, 85748 Garching bei M\"unchen,
  Germany\\$^{5}$National Radio Astronomy Observatory, 520 Edgemont
  Road, Charlottesville, VA 22903, USA\\$^{6}$Mullard Space Science
  Laboratory, University College London, Holmbury St Mary, Dorking,
  Surrey RH5 6NT, UK}
        
\begin{document}



\maketitle

\label{firstpage}

\begin{abstract}
We use new mid-infrared (mid-IR) photometry from the {\em Spitzer
  Space Telescope} to study the relations between low-frequency radio
luminosity density $L_{\nu \rm 151MHz}$, mid-IR ($\rm 12\,\mu$m
rest-frame) luminosity $\nu L_{\nu \rm 12\mu m}$, and
optical-emission-line ([OII]) luminosity $L_{\rm [OII]}$, for a
complete sample of $\rm z\sim1$ radio galaxies from the 3CRR, 6CE,
6C*, 7CRS and TOOT00 surveys. The narrow redshift span of our sample
($0.9<z<1.1$) means that it is unbiased to evolutionary effects. We
find evidence that these three quantities are positively
correlated. The scaling between $\nu L_{\nu \rm 12\mu m}$ and $L_{\rm
  [OII]}$ is similar to that seen in other AGN samples, consistent
with both $\nu L_{\nu \rm 12\mu m}$ and $L_{\rm [OII]}$ tracing
accretion rate. We show that the positive correlation between $\nu
L_{\nu \rm 12\mu m}$ and $L_{\nu \rm 151MHz}$ implies that there is a
genuine lack of objects with low values of $\nu L_{\nu \rm 12\mu m}$
at high values of $L_{\nu \rm 151MHz}$. Given that $\nu L_{\nu \rm
  12\mu m}$ traces accretion rate, while $L_{\nu \rm 151MHz}$ traces
jet power, this can be understood in terms of a minimum accretion rate
being necessary to produce a given jet power. This implies that there
is a maximum efficiency with which accreted energy can be chanelled
into jet power and that this efficiency is of order unity.
\end{abstract}


\section{Introduction}

The currently most-accepted scheme for the structure of an active
galactic nuclei (AGN), the unified model, consists of a supermassive
  black hole (SMBH), surrounded by an accretion disc and, in the same
  plane at larger distances, an axisymmetric distribution of gas and
  dust, usually referred to as a torus. Within the inner
  region of the obscuring torus lies an ensemble of highly ionized and
  dense gas clouds, where broad emission lines are produced, the broad
  line region (BLR). Outside the torus there is a region of
  low-ionization less dense gas clouds, the narrow line region (NLR)
  (e.g. Antonucci 1993, Urry \& Padovani 1995).

The dust present in the putative obscuring torus blocks most radiation
from the inner photoionizing source at optical, UV and soft X-ray
wavelengths, but it is responsible for most of the AGN emission at
mid-IR frequencies (e.g. \citealt{1985ApJ...297..621A},
\citealt{1990MNRAS.242P...4B}, \citealt{1998ApJ...498..570D},
\citealt{2007A&A...474..837T}). The dust in the torus absorbs and
re-radiates the emission from the core at these frequencies. Thus,
mid-IR observations, for example those probing rest-frame wavelengths
around 12\,$\mu$m, indirectly reveal radiation from regions close to
the active nucleus. This is supported by observations that show the
mid- to far-IR continuum emitted by powerful radio galaxies to be
mainly due to AGN heating of circumnuclear dust (e.g. Dicken et
al. 2009).

According to the most widely accepted model for the archetype
structure of an AGN, in the accretion disc, and potentially in regions
connected to the event horizon of the black hole itself, magnetized
inflowing material enables processes capable of driving two opposed
collimated outflows, known as the jets
(e.g. \citealt{1973MNRAS.164..243L}, \citealt{1979ApJ...232...34B},
\citealt{2001Sci...291...84M}). These jets push magnetised material
out of the central region and feed lobes. The acceleration of
ultrarelativistic charged particles produces synchrotron radiation,
visible mainly at radio frequencies
(e.g. \citealt{1974ApJ...192..261J}). Therefore, the radio luminosity
due to jets is inherently related to the central engine that powers
the AGN.

One of the most prominent features in the optical spectrum of an AGN
is the [OII] emission line. It is produced in a very low density
ionized medium, such as the NLR, when thermal photoelectrons collide
with oxygen atoms and excite their lower energy levels. Even though
de-excitation through downward radiation has a very low probability,
in such a low density medium, collisions are much more sparse, so the
downward radiation results in the emission of this forbidden-line.
Since the BLR can be obscured by the dusty torus, depending on the
galaxy's orientation, and the NLR is always visible, [OII] is a good
tracer of the underlying continuum emission in an AGN
(e.g. \citealt{1993Natur.362..326H}), although other narrow lines,
e.g. [OIII], may be better \citep{1998MNRAS.297L..39S}.

It is now well established that there is a correlation between the
luminosity of the NLR and the radio luminosity for radio-loud
objects and that this relation is independent of redshift
(e.g. \citealt{1991Natur.349..138R}, \citealt{1999MNRAS.309.1017W}). This
supports the idea that the physical process behind radio jets is
intrinsically linked to the source of the narrow lines via a common
central engine (an accreting SMBH).

In this paper we explore these ideas further by showing how rest-frame
$\rm 12\,\mu m$ luminosity, $\nu L_{\nu \rm 12\mu m}$, as a probe of
warm dust emission, correlates with $L_{\nu \rm 151MHz}$, a tracer of
jet power, and with $L_{\rm [OII]}$, which we assume tracks the
underlying continuum emission of the AGN.

Throughout this paper we adopt the following values for the
cosmological parameters: $\rm H_0=70\,km\,s^{-1}\,Mpc^{-1}$, $\rm
\Omega_M=0.7$ and $\rm \Omega_\Lambda=0.3$. We use the convention
$S_{\nu}\propto\rm\nu^{-\alpha}$.

\section{Sample definition}

\begin{figure}
  \begin{center}
    \includegraphics[width=1.0\columnwidth]{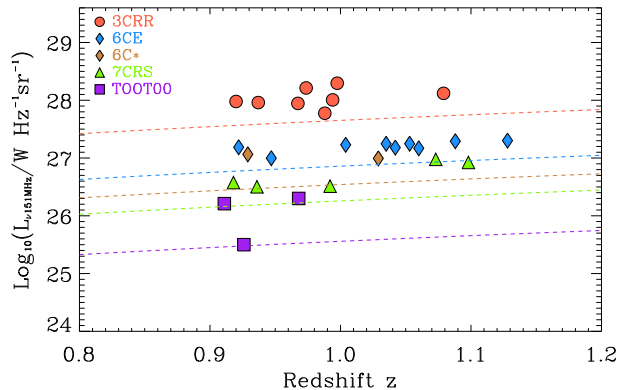}
  \end{center}
  \caption{Radio luminosity density at rest-frame 151\,MHz versus
    redshift for the 3CRR (red circles), 6CE (blue diamonds), 6C*
    (orange diamonds), 7CRS (green triangles) and TOOT00 (purple
    squares) radio sources with $0.9<z<1.1$. The dashed lines show the
    loci corresponding to the radio flux density limit of the
    different surveys assuming a spectral index, $\alpha= 0.8$.}
  \label{fig:l151_z}
\end{figure}

We select all of the narrow-line radio galaxies from the complete
low-frequency selected radio samples 3CRR \citep{1983MNRAS.204..151L},
6CE \citep{1997MNRAS.291..593E,2001MNRAS.322..523R}, 6C* (
\citealt{1998MNRAS.295..265B,2001MNRAS.326.1563J}a,b), 7CRS
\citep{1999MNRAS.308.1096L,2003MNRAS.339..173W}, and TOOT00
\citep{2003NewAR..47..373H,Vardoulaki} surveys by requiring a narrow
redshift span of $0.9<z<1.1$. The low selection frequency of these
samples (either 178 or 151~MHz) ensures that they are selected based
on their optically thin lobe emission and thus in an orientation
independent way. The narrow-line selection ensures that we are not
including Blazars or type-I AGN, although 3C343 has since been
classified as a QSO (Cleary et al. 2007). Assembling together galaxies
from the different radio surveys, allows a wide range of radio
luminosities ($25 <\log_{10}(L_{\nu \rm 151MHz}/\rm
\,W\,Hz^{-1}\,sr^{-1})<29$) to be covered. At the same time, such a
narrow width in redshift allows us to study correlations independent
of possible evolutionary effects and any issues with assumed
k-corrections. In the local Universe there is not enough cosmic volume
to gather a reasonable number of luminous sources, whereas at high
redshifts there are no statistically complete samples for the low
flux limits that distant luminous sources require to be observed,
therefore $0.9 < z < 1.1$ was chosen as a reasonable compromise. The
span of our sample in the redshift--radio luminosity plane is shown in
Figure~\ref{fig:l151_z}.

Most of the radio galaxies in our sample are classified as FRIIs,
i.e. they are brightest at the edges of the radio structure. The only
exception is 3C343, which is core dominated and has since been
identified as a quasar \citep{2007ApJ...660..117C}, and 6CE1217+37 and
5C7.242, both of which could be classified as either an FRI or FRII.

\section{Observations}

The observations presented in this paper were made with the Multiband
Imaging Photometer for Spitzer (MIPS) camera on the Spitzer Space
Telescope which provides long-wavelength capability, under the
programme ID30344 (PI. Jarvis).  Our observations were carried out
between August 2006 and August 2007. Our objects are relatively bright
and short exposures were sufficient. For the quasars we observed for
one cycle using 7 jitters of 10 seconds, resulting in a total exposure
time of 70~seconds on source.  For the radio galaxies we adopted a
slightly different strategy; the 3CRR sources were observed with the
same strategy as the quasars (i.e. 70 second total exposure time),
while the less radio luminous objects from the 6CE, 7CRS and TOOT
samples were observed for two 7-jitter cycles of 10 second per jitter,
resulting in 140~seconds on source, to ensure significant detections,
as previous observations have shown there to be a correlation between
host galaxy mass/luminosity and the radio luminosity for radio
galaxies (e.g. Willott et al. 2003; McLure et al. 2004).  Some of the
3C radio galaxies already had adequate data in the archive, for MIPS
these include 3C22, 3C184, 3C280, 3C268.1, 3C289 and 3C343 (ID74; PI
Houck) and 3C356 (ID3329; PI Stern).  The data reduction was performed
using the standard pipeline version S15.0.5. We measured the 24$\mu$m
flux densities by using an aperture diameter of 6.12 pixels and a 1.61
aperture correction, to match the procedure used by the MIPS
instrument team to derive calibration factors from standard star
observations.  Table \ref{table:summary} presents a summary of the
characteristics of the selected sources.

\begin{table*}
\caption{\textbf{Column 1} gives the name of the object;
  \textbf{Column 2} gives redshift; \textbf{Column 3} gives
  low-frequency radio luminosity; \textbf{Column 4} gives the
  rest-frame 12\,$\mu$m luminosity; \textbf{Column 5} gives [OII]
  emission line luminosity; and \textbf{Column 6} gives a reference
  for [OII] flux when available, or luminosity. References are
  identified as follows: B00 - Best et al 2000; I02 - Inskip et
  al. 2002a; J01 - Jarvis et al. 2001; L96 - Lawrence et al. 1996; M88
  - McCarthy 1988; M95 - McCarthy et al. 1995; R01 - Rawlings et
  al. 2001; W02 - Willott et al. 2002; W03 - Willott et al. 2003; V09
  - Vardoulaki et al. 2009; \textbf{Column 7} gives the
    optical classification of the sources. ``G'' stands for galaxy,
    ``QSO'' for quasar, a ``?'' shows uncertainty. } \centering
\begin{tabular}{l c c r c c c}
\hline\hline
\multicolumn{1}{c}{Object} & Redshift & $\log_{10}(L_{\nu \rm 151MHz}/$ & \multicolumn{1}{c}{$\log_{10}(\nu L_{\nu \rm 12\mu m}/$} & $\log _{10}(L_{\rm[OII]}/$  & Ref ([OII]) & Opt. Class.\\[0.5ex]
 & & $\rm W\,Hz^{-1}\,sr^{-1}$) & \multicolumn{1}{c}{$\rm W)$} & $\rm W$) & & \\
\multicolumn{1}{c}{(1)} & (2) & (3) & \multicolumn{1}{c}{(4)} & (5) & (6) & (7)\\
\hline
3C280        & 0.997 & 28.29 & 38.78$\pm0.01$ & 36.04 & B00 & G\\
3C268.1      & 0.974 & 28.21 & 37.76$\pm0.06$ & 35.54 & M88 & G\\
3C356        & 1.079 & 28.12 & 38.51$\pm0.02$ & 35.68 & B00 & G\\
3C184        & 0.994 & 28.01 & 37.68$\pm0.11$ & 35.89 & M95 & G\\
3C175.1      & 0.920 & 27.98 & 37.65$\pm0.09$ & 35.93 & M88 & G\\
3C22         & 0.937 & 27.96 & 38.88$\pm0.01$ & 35.95 & B00 & G\\
3C289        & 0.967 & 27.95 & 38.34$\pm0.02$ & 35.57 & B00 & G\\
3C343        & 0.988 & 27.78 & 38.67$\pm0.01$ & 35.00 & L96 & QSO\\
6CE1256+3648 & 1.128 & 27.30 & 38.08$\pm0.04$ & 35.17 & I02 & G\\
6CE1217+3645 & 1.088 & 27.29 & 37.40$\pm0.15$ & 34.52 & I02 & G\\
6CE1017+3712 & 1.053 & 27.25 & 37.93$\pm0.05$ & 35.61 & I02 & G\\
6CE0943+3958 & 1.035 & 27.25 & 38.15$\pm0.03$ & 35.45 &I02 & G\\
6CE1257+3633 & 1.004 & 27.23 & 37.75$\pm0.06$ & 35.07 & I02 & G\\
6CE1019+3924 & 0.922 & 27.19 & 37.25$\pm0.26$ & 35.02 & I02 & G\\
6CE1011+3632 & 1.042 & 27.18 & 37.98$\pm0.05$ & 34.86 & I02 & G\\
6CE1129+3710 & 1.060 & 27.17 & 37.81$\pm0.06$ & 35.49 & I02 & G\\
6C*0128+394  & 0.929 & 27.06 & 36.87$\pm0.33$ & 34.46 & J01 & G\\
6CE1212+3805 & 0.947 & 27.00 & 37.17$\pm0.16$ & 35.14 & R01 & G\\
6C*0133+486  & 1.029 & 26.99 & 36.76$\pm0.70$ & 35.00 & J01 & G\\
5C6.24  (7CRS) & 1.073 & 26.98 & 37.75$\pm0.07$ & 35.51 & W02 & G\\
5C7.23  (7CRS) & 1.098 & 26.92 & 37.71$\pm0.11$ & 35.12 & W02 & G\\
5C7.82  (7CRS) & 0.918 & 26.57 & 37.35$\pm0.17$ & 34.80 & W02 & G\\
5C7.242 (7CRS) & 0.992 & 26.51 & 38.11$\pm0.04$ & 34.66 & W02 & G\\
5C7.17  (7CRS) & 0.936 & 26.50 & 37.88$\pm0.06$ & 35.25 & W03 & G\\
TOOT00\_1267     & 0.968 & 26.30 & 37.81$\pm0.05$  & 34.66 & V09 & G?\\
TOOT00\_1140     & 0.911 & 26.21 & 36.85$\pm0.29$ & 33.92 & V09 & G\\
TOOT00\_1066     & 0.926 & 25.50 & 37.21$\pm0.13$ & 34.69 & V09 & G\\
\hline
\end{tabular}
\label{table:summary}
\end{table*}

The $\rm \nu L_{\nu 12\mu m}$ values were calculated from Spitzer/MIPS
observations at 24\,$\rm \mu m$ using the textbook equation
(e.g. \citealt{1999coph.book.....P}):

\begin{equation}
L_{\nu}([1+z]\nu_{0})=\frac{4\pi D_{\rm L}^{2}S_{\nu}(\nu_{0})}{(1+z)},
\end{equation}
where $S_{\nu}(\nu_{0})$ is the flux density at the observed
frequency, $L_{\nu}([1+z]\nu_{0})$ is the luminosity density at the
emitted frequency, and $D_{\rm L}$ is the luminosity distance. Given
that $\nu_{12\mu m}$ is very close to $\nu_{24\mu m /(1+z)}$ for this
particular $z\sim 1$ sample, and that at rest-frame 12\,$\mu$m
the spectrum in $\nu L_{\nu}$ is quite flat, we approximate the
mid-infrared spectrum in this region to a power law with spectral index $\rm
\alpha_{12\mu m}=1$ in our calculations.

The [OII] emission line flux, $S_{\rm [OII]}$, values were taken from
the literature and converted to luminosity (see
Table~\ref{table:summary} for references), or when only $L_{\rm
  [OII]}$ values were available, these were converted to our adopted
cosmology. Despite the fact that the [OIII]
  narrow-emission-line may be a better tracer of the underlying
  continuum (Simpson 1998), at $z=1$ the [OII] emission-line lies in
  the optical part of the spectrum and is therefore available for our
  whole sample, whereas the [OIII] line is redshifted into the
  near-infrared.

\section{Results}

\subsection{Relation between $L_{\nu \rm 151MHz}$ and $L_{\rm [OII]}$}

\begin{figure}
  \begin{center}
    \includegraphics[width=1.0\columnwidth]{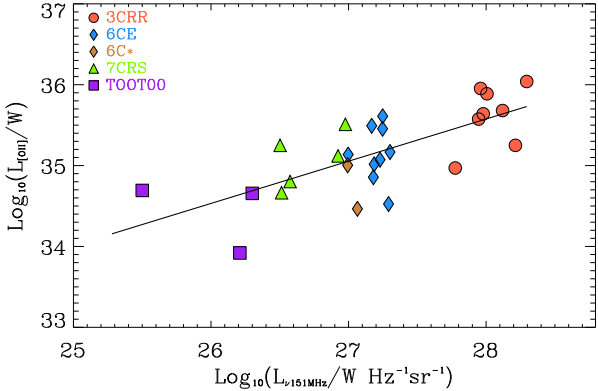}
  \end{center}
  \caption{Luminosity of [OII], $L_{\rm [OII]}$, versus Luminosity at
    151MHz, $L_{\nu \rm 151MHz}$. Symbols are as in
      Figure~\ref{fig:l151_z}. The solid black line is the best fit
    line for all objects and its equation is given by $\log_{10}(
    L_{\rm [OII]})= (0.52\pm 0.10)\log_{10}(L_{\nu \rm
      151MHz})+21.0\pm 2.7$. The Spearman correlation coefficient is
    $\rho=0.68$ with a Spearman rank correlation probability of 99.9\%}
  \label{fig:l151_loii}
\end{figure}

Figure \ref{fig:l151_loii} presents the [OII] line luminosity against
low-frequency radio luminosity for the selected sample of radio
galaxies. It shows that there is a positive correlation between these
two quantities. The best-fitting relation is $L_{\rm [OII]}\propto
L_{\nu \rm 151MHz}^{0.52\pm0.10}$, also plotted on
Figure~\ref{fig:l151_loii}. The slope and intersect of this relation
and the errors associated with them were found by applying a Bayesian
linear regression (e.g. \citealt{Sivia}). The Spearman correlation
coefficient is $\rm \rho=0.68$, with a 99.9\% probability that the
null hypothesis, i.e. that these two quantities are not correlated,
can be dismissed.

This confirms previous results (Rawlings et al. 1989, Rawlings $\&$
Saunders 1991, McCarthy 1993, Willott et al. 1999; Jarvis et al. 2001a) that found narrow
emission-line luminosities and radio luminosities to be positively
correlated. Willott et al. (1999) performed similar studies on sources
from the 3CRR and 7CRS surveys that span a wide range of redshifts and
found a similar relation: $L_{\rm [OII]}\propto L_{\nu \rm
  151MHz}^{\beta}$, with $\beta\sim0.8$ with significant uncertainties
in this value as it appears to vary with the precise sub-set of
objects under investigation.

The previous results typically sample objects with
the highest radio luminosities at each redshift. Thus, due to volume
and evolutionary effects, objects in the local Universe will have
lower luminosities whereas sources at high redshifts will have higher
luminosities, spreading the objects over a wide range of
radio-luminosities. The complicated selection function
makes it unsurprising that our sample, cleaner in the important sense
that all objects have the same redshift, has a slightly different
slope. In addition, in a flux limited sample,
  luminosity functions dictate that objects from the faint end will
  preferably be sampled over objects from the bright end, enforcing a
  steeper correlation compared to a sample that spreads over a
  wider range of radio luminosities.

\subsection{Relation between $L_{\nu \rm 12\mu m}$ and $L_{\rm [OII]}$}

\begin{figure}
  \begin{center}
    \includegraphics[width=1.0\columnwidth]{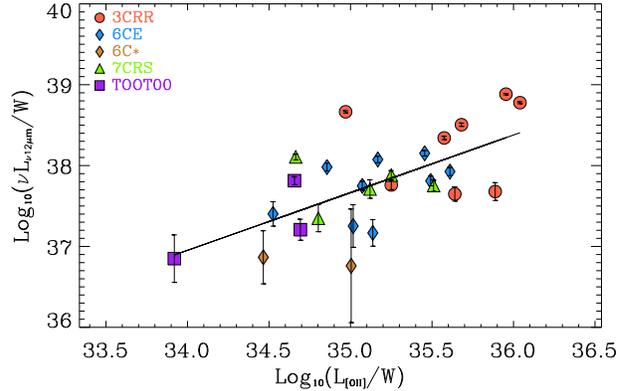}
  \end{center}
  \caption{Luminosity at 12\,$\mu$m rest-frame, $\nu L_{\nu \rm 12\mu
      m}$ versus Luminosity of [OII] emission line, $L_{\rm
      [OII]}$. Symbols are as in Figure~\ref{fig:l151_z}. The
    solid black line is the best fit line for all objects and its
    equation is given by $\log_{10}(\nu L_{\nu \rm 12\mu m})= (0.71\pm
    0.17)\log_{10}(L_{\rm [OII]})+12.7\pm 5.8$. The Spearman
    correlation coefficient is $\rho=0.54$ with a Spearman
        rank correlation probability of 99.7\% }
  \label{fig:loii_l12}
\end{figure}

Figure~\ref{fig:loii_l12} shows the distribution of rest-frame mid-IR
(12\,$\mu$m) luminosity of the selected sample of galaxies with
relation to [OII] luminosity. We find that these quantities are
strongly correlated with the best-fitting solution for this relation
$L_{\rm [OII]}\propto \nu L_{\nu \rm 12\mu m}^{0.71\pm0.17}$. The
Spearman correlation coefficient is $\rho=0.54$ with a Spearman rank
correlation probability of 99.7\%.

A previous study conducted by Dicken et al (2009), using a different
sample of powerful radio sources ($0.05<z<0.7$), also showed a tight
correlation exists between the [OIII] luminosity and mid- to
far-infrared (24\,$\mu$m and 70\,$\mu$m) luminosities, with the same
slope value of approximately 0.71. The authors propose that since
[OIII] luminositiy is a good indicator of the intrinsic AGN power
(e.g. \citealt{1998MNRAS.297L..39S}), these correlations indicate that
mid- to far-infrared luminosities must also be linked to the power of
the active core.

Following a similar argument, our correlation between [OII] and
12\,$\mu$m luminosities is consistent with a view in which the central
AGN is the source of radiation that excites both the NLR, ionizing its
gas clouds, and the circumnuclear dust of the inner torus area, where
it gets absorbed and re-emitted.

Although this correlation could naively be expected to lie closer to a
proportionality, some factors are expected to cause some differences
in the way 12\,$\mu$m and [OII] emission trace the central
emission. For instance, while emission at 12\,$\mu$m is powered by
optical-UV radiation from the accretion disc, the [OII] emission line
is excited by ionizing radiation from the disc, and therefore
variations of the ionization parameter with luminosity might induce a
trend in the relationship. In addition, the 12\,$\mu$m
emission is radiated from a region closer to the
nucleus and thus it might not be as isotropic as [OII] emission
(see e.g. \citealt{2007ApJ...660..117C}). Also, in light of
the receding torus model, the proportion between mid-IR and [OII]
luminosities might vary with luminosity (as previously suggested by
Dicken et al. 2009). That is, in more powerful AGN, the maximum radius
at which dust sublimates is further away from the central engine,
enlarging the opening angle of the ionizing cone and thus increasing
the [OII] emission \citep[e.g.][]{1991MNRAS.252..586L}.

\subsection{Relation between $L_{\nu \rm 151MHz}$ and $\nu L_{\nu \rm 12\mu m}$}

\begin{figure*}
  \begin{center}
    \includegraphics[width=2.0\columnwidth]{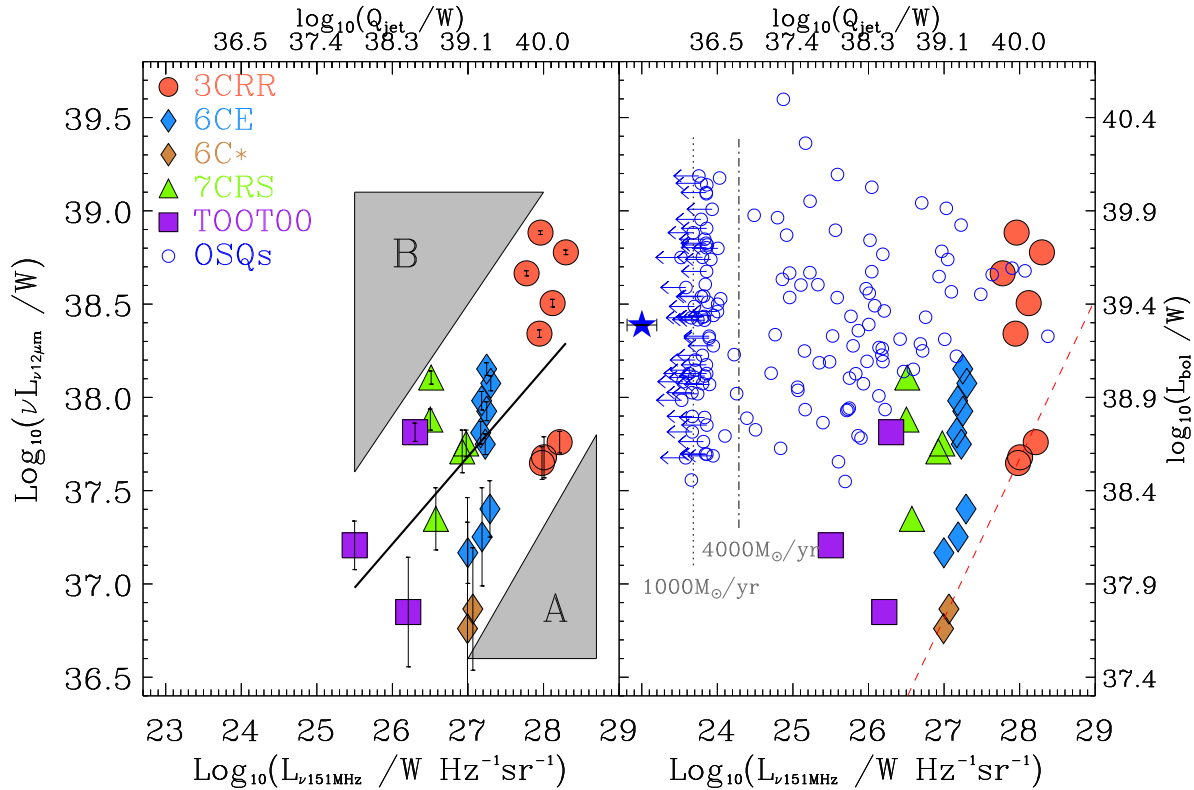}
  \end{center}
  \caption{\textbf{Left:} Mid-IR (12\,$\mu$m) rest-frame luminosity,
    $\nu L_{\nu\rm 12\mu m}$, versus low-frequency radio luminosity,
    $L_{\nu \rm 151MHz}$. Symbols are as in
    Figure~\ref{fig:l151_z}. The solid black line is the best fit line
    given by $\log_{10}(\nu L_{\nu \rm 12\mu m})= (0.47\pm
    0.13)\log_{10}(L_{\nu \rm 151MHz})+25.0\pm 3.6$. The Spearman
    correlation coefficient is $\rho=0.48$ with a Spearman rank
    correlation probability of 98.8\%. \textbf{Right:} Same plot as on
    the left overlaid with a sample of bright optically selected
    quasars (OSQs) with $0.9<z<1.1$ (blue cicles) from Falder et
    al. (2010). The top x-axes show the corresponding jet power,
    estimated using $Q_{\rm jet}\simeq 3\times 10^{38}f^{3/2}( L_{\nu
      \rm 151MHz} / 10^{28} )^{6/7} \rm W $ (Willott et al. 1999), the
    left-hand y-axis shows bolometric luminosity, estimated using
    $L_{\rm bol}\simeq8.5\times \nu L_{\nu 12 \mu \rm m}$
    \citep{2006ApJS..166..470R}. The fact that these objects all lie
    above $\log_{10}(\nu L_{\nu \rm 12\mu m})\sim 37.5$ is due to the
    minimum infrared luminosity corresponding to the optical
    sensitivity limit of SDSS.  The vertical dot-dashed line shows the
    radio luminosity emitted by a starburst forming massive stars
    ($M\geq 5\rm \,M_{\sun}$) at a rate of $1000\,\rm M_{\sun}/yr$
    \citep{1992ARA&A..30..575C}, equivalent to a total star formation
    rate of $\sim 4000\,\rm M_{\sun}/yr$ assuming a Salpeter initial
    mass function. The vertical dotted line shows the radio luminosity
    emitted by a starburst forming massive stars at a rate of
    $250\,\rm M_{\sun}/yr$, equivalent to a total star formation rate
    of $\sim 1000\,\rm M_{\sun}/yr$ assuming a Salpeter initial mass
    function.  All quasars with flux $<2\sigma$ are represented as
    limits. The blue filled star represents the average of the stack
    of all the OSQs with limits. The red dashed line represents the
    maximum jet efficiency, found for $\eta\sim2.5$.  }
  \label{fig:l151_l12}
\end{figure*}

The left panel of Figure~\ref{fig:l151_l12} shows the correlation of
the mid-IR (12\,$\mu$m) rest-frame luminosity density against
low-frequency radio luminosity density for the selected sample of
radio galaxies. It is clear that more powerful radio galaxies exhibit
more luminous mid-IR emission than less powerful ones. The
best-fitting model for this relation is $\nu L_{\nu \rm 12\mu
  m}\propto L_{\rm 151MHz}^{0.47\pm0.13}$. The Spearman correlation
coefficient is $\rho=0.48$ with a Spearman rank correlation
  probability of 98.8\%.

Cleary et al. (2007) found a similarly strong 
correlation between 15\,$\mu$m luminosity and 178\,MHz total radio
luminosity for the $0.4<z<1.2$ 3CRR sample, further confirmed by
Hardcastle et al. (2009) for the $z<1.2$ sources of the same survey
(see also Ogle et al. 2006). By further narrowing the redshift range
we show that the correlation for radio-loud AGN is still significant
and it is therefore not due to an underlying correlation between
redshift and luminosity.

Figure~\ref{fig:l151_l12} shows that there are two `gaps' of objects
in our sample: objects with high $L_{\nu \rm 151MHz}$ and low $\nu
L_{\nu \rm 12\mu m}$, and objects with high $\nu L_{\nu \rm 12\mu m}$
and low $L_{\nu \rm 151MHz}$, respectively represented as regions
``A'' and ``B'' on the plot.  We believe that region A is not an
artifact of selection but a real gap. The justification for this lies
on the following points: (1) The luminosity function of galaxies at
all wavelengths shows that fainter objects always outnumber brighter
objects and therefore we can always expect to observe more fainter
objects for a given wavelength range. (2) Spitzer MIPS 24\,$\mu$m is
sensitive enough to detect objects in that range of rest-frame
12\,$\mu$m luminosities but finds only objects much brighter than the
24\,$\mu$m flux limit. (3) 3CRR, for instance, is a full sky survey
and therefore the lack of objects with lower values of $\nu L_{\nu \rm
  12\mu m}$ is not due to insufficient area surveyed. (4) Similarly,
it cannot be due to cosmic volume, as the volume surveyed is enough to
detect bright objects at rest-frame 12\,$\mu$m. We note that Bl
  Lac objects could effectively populate region A, however the
  relativistic beaming thought to occur in Bl Lacs means that their
  inferred radio luminosity is not a fair indicator of the total radio
  emission of the object and upon accounting for this the flux would
  lie to the left of region A.

The situation with region B is fundamentally different. Our sample is
radio selected as opposed to infrared selected so we expect objects
with fainter 12\,$\mu$m luminosities to outnumber those with high $\nu
L_{\nu \rm 12\mu m}$ (see point (1) above). The surveys deeper than
3CRR cover decreasingly smaller sky areas as their flux density limit
decreases, thus reducing the likelihood of detecting objects with high
$\nu L_{\nu \rm 12\mu m}$ (see points (3) and (4) above). For these
reasons, the gap B of objects could be the result of selection
effects.

It is therefore useful to include objects that lie outside the regime
of high-$S_{\rm 151MHz}$-selected samples, as well as objects known to
be brighter in IR.  For this purpose, we added a sample of
bright optically selected quasars at $0.9\leq z\leq 1.1$ with both
radio quiet (RQQs) and radio loud quasars (RLQs), shown on
  the right panel of Figure~\ref{fig:l151_l12}. The optically bright
quasar sample was selected from the Sloan Digital Sky Survey (SDSS)
following the QSO selection criteria of
\citet{2004ApJS..155..257R}. To pick up RQQs and RLQs, they were
cross-referenced with the Very Large Array (VLA) Faint Images of the
Radio Sky at Twenty-cm (FIRST) survey \citep{1995ApJ...450..559B} and
the Westerbork Northern Sky Survey (WENSS;
\citealt{1997A&AS..124..259R}). The RQQs were required to be
undetected by the FIRST survey at the 5$\sigma$ level, whereas the
RLQs were hand-picked to have a low-frequency WENSS (325\,MHz) flux
density greater than 5$\sigma$ the limit of the survey. More details
about the sample selection can be found in Jarvis et al. (\textit{in
  prep.}) and \citet{2010MNRAS.405..347F}.

The full sample of objects scatter on top and to the left of our
initial sample of radio galaxies. However, gap A is still present,
with the radio galaxies from the original sample delimiting a lower
envelope to the rest of the sample. The objects fill up gap B, indeed
showing it is the result of selection effects.

Some of the objects were not detected with the FIRST survey at a
$2\sigma$ level and these are represented as limits. To determine
where approximately these non-detections lie, we have used a stacking
analysis (yielding statistical detections) to determine that they do
not lie more than about an order of magnitude away from the FIRST flux
density limit. We stacked together the radio images from FIRST of all
the non-detections at their known position and obtained a detection at
a $2.2\sigma$ level. The blue filled star represents the average
luminosities of the stack of all the quasars with limits.

\subsection{Physical Interpretation}

Radio luminosity traces the jet power and, in particular for FRIIs,
Willott et al (1999) has shown that the jet power, $Q_{\rm jet}$,
scales with $L_{\nu \rm 151MHz}$ through the relation $Q_{\rm
  jet}\simeq 3\times 10^{38}f^{3/2}( L_{\nu \rm 151MHz} / 10^{28})^{6/7} \rm W $,
where $1\leq f\leq 20$ represents several uncertainties associated
with estimating $Q_{\rm jet}$ from $L_{\nu 151MHz}$. We chose $f=10$ as
this is the expectation value of a flat prior in natural space. 

At the same time, the re-radiated mid-IR emission, probes the
radiation emitted by the accretion disc and hence the bolometric
 luminosity $L_{\rm bol}$. In turn,  $L_{\rm bol}$ is directly linked
to the accretion rate, $\dot{M}$, by $L_{\rm bol}=\epsilon \dot{M}
c^2$, where $\epsilon$ is the radiative efficiency of the accretion
process. Therefore, a significant correlation between $L_{\nu \rm
  151MHz}$ and $\nu L_{\nu \rm 12\mu m}$ over a wide range of radio
luminosities indicates that there is physical relationship between jet
power and accretion rate \citep{1991Natur.349..138R}.

Indeed, if we assume the radiative efficiency to be approximately constant,
with a value of $\epsilon=0.1$ (consistent with theoretical
simulations, such as \citealt{2008MNRAS.390...21B}, and observational
constrains, for example \citealt{2009ApJ...692..964M}), then the bolometric
luminosity is proportional to the accretion rate, $L_{\rm bol} \propto
\dot{M}$.

The fact that there is a genuine gap (gap A) of objects with high
$L_{\nu \rm 151MHz}$ and low $\nu L_{\nu \rm 12\mu m}$ reveals that
to achieve a given jet power, there is a minimum necessary accretion rate.
If the jet power can be parameterised as a function of accretion rate by
$Q_{\rm jet}=\eta \dot{M}c^{2}$, with a jet efficiency term
$\eta$, then, for these powerful radio galaxies, a minimum accretion
rate implies a maximum jet efficiency. 

To quantify how much this maximum efficiency would be, we can use
$\eta = \epsilon~Q_{\rm jet}/L_{\rm bol}$ from above and apply it to
the galaxies that form the lower envelope of the data points just
above gap A. This yields $Q_{\rm jet}/L_{\rm bol}\sim
10^{39.1}/10^{37.7}\approx 25.1$, which, with the canonical value of
$\epsilon=0.1$, gives $\eta \approx 2.5$ (see red dashed-line in
Figure\,\ref{fig:l151_l12}), a rather high jet efficiency. On the same
line, \citet{2010arXiv1007.1227M}, have found that for some sources
$Q_{\rm jet}>\dot{M}c^{2}$, and thus $\eta>1$. The value found is
dependent on the uncertaintes of estimating $Q_{\rm jet}$ from $L_{\rm
  151MHz}$ and $L_{\rm bol}$ using bolometric corrections. Therefore,
parameters like $f$ can significantly alter the value of $\eta$. For
instance, for $f=1$ to $f=20$, the maximim jet efficiency can span the
interval $0.08\lesssim \eta \lesssim
7.9$. \citet{2006ApJ...651L..17P,2007MNRAS.374L..10P} also finds that
FRIIs can exist in a state with a high time-averaged ratio of $Q_{\rm
  jet}/L_{\rm bol}\sim20$ or even higher, and finds theoretical
solutions with a maximum time averaged ratio of $Q_{\rm jet}/L_{\rm
  bol}\sim10$. As discussed in Punsly (2007), the value of maximum jet
efficiency, or, equivalently, $(Q_{\rm jet}/L_{\rm bol})_{\rm max}$,
has important theoretical repercussions in terms of whether the jets
of FRIIs are powered by the accretion disk or the black hole itself.

There is some discussion as to what is the most appropriate value for
$f$. Cavagnolo et al. (2010) use Chandra X-ray observations to compute
the energy stored in X-ray cavities and estimate the jet power. They
then use radio data to investigate the scaling between jet power and
radio luminosity, and find a relation that has the same slope as that
of Willott et al. (1999), but where an $f=1$ gives a relation that is
two orders of magnitude below their normalisation. An $f\sim 20$ seems
thus more in agreement with these studies. Similarly
Mart\'inez-Sansigre \& Rawlings (2010) find that a
minimum of $f=20$ is required to fit the local radio luminosity
function of galaxies. On the other hand, some studies (e.g. Croston et
al. 2003, 2004) suggest that the value of $f$ should be high for FRIs
and low for FRIIs, $f\sim$ 10-20, and low for FRIIs, $f\sim$ 1-2.

The regime at which not all radio
galaxies are powerful accretors occurs at a lower radio luminosity
($<5\times 10^{26}\,\rm W\,Hz^{-1}\,sr^{-1}$), as can be seen in
Figure~1 of \citet{2006ApJ...647..161O}.

Given that $\nu L_{\nu \rm 12\mu m}$ is a proxy for bolometric
luminosity and hence accretion rate, and $L_{\nu \rm 151MHz}$ traces
the jet power produced by the SMBH, the very large scatter observed in
the right panel of Figure~\ref{fig:l151_l12} indicates that galaxies
with similar accretion rates can produce powerful or weaker
jets. Therefore the scatter is probably due to a continuous range of
efficiencies with which AGN produce jets. The factors that rule this
efficiency are still unclear, but black hole spin might be a dominant
mechanism (e.g. see \citealt{2007ApJ...658..815S} and references
within). Spin is indeed an appealing mechanism for it allows nominal
efficiencies $\gtrsim 1$, since some of the jet power originates from
rotation of the black hole, rather than from the accreted
mass\footnote{Note, however, that high jet efficiencies involving high
  spins would also increase the radiative efficiency to values as high
  as $\sim 0.3$ (\citealt{1970Natur.226...64B},
  \citealt{1974ApJ...191..507T}).}
(e.g. \citealt{1984ARA&A..22..471R}).

The fact that [OII] line emission luminosity is positively correlated
with $12\,\rm \mu m$ rest-frame luminosity implies that both
quantities trace accretion rate.  As $L_{\rm [OII]}$ is directly
linked to the radiative photoionizing power, which is itself a
fraction of the bolometric luminosity, $L_{\rm bol}$, $L_{\rm [OII]}$
is therefore also a tracer of accretion rate. 

The relation between $L_{\rm [OII]}$ and $L_{\nu \rm 151MHz}$ has
similar implications as the relation between $\nu L_{\nu \rm 12\mu m}$
and $L_{\nu \rm 151MHz}$. In fact we expect that with the
  inclusion of weaker radio sources, the relation between $L_{\nu \rm
    151MHz}$ and $L_{\rm [OII]}$ would become less tight and probably
  convert into an envelope, in a similar way to what happens with the
  radio-loud objects for the $L_{\nu\rm 151MHz}$-$\nu L_{\nu \rm 12\mu m}$
  relation. Some studies of weaker radio sources (e.g. Best et
  al. 2005b; Shabala et al. 2008) indeed suggest that there is a
  trend for radio-quiet sources to scatter beyond the tight
  correlation observed for our sample of radio loud sources. This
  could also be related to a different accretion mechanism for the
  lower power radio sources (e.g. Hardcastle et al. 2006).

Some attention should be drawn to the fact that radio luminosity is
not a perfect tracer of the kinetic jet power as it also depends on
environmental effects irrespective of the jet power. For instance, the
characteristics of the jet environment may play a significant role in
how efficiently the jet power gets converted into radio
luminosity. Especially in the richer environments inhabited by the
strongest radio sources, a dense environment might be responsible for
boosted radio luminosities at a given jet power (e.g. Barthel $\&$
Arnaud 1996, Falder et al 2010). Moreover, the low frequency radio
emission is radiated over scales that are orders of magnitude larger
in extent than the torus and hence slower to react to any changes in
the central emission than the circumnuclear dust or the NLR.

Another caveat is that superimposed on the spectrum at 10\,$\mu$m
there is a silicate feature due to the Si-O streching mode of
amorphous silicate grains of dust
(e.g. \citealt{2008NewAR..52..274E}). This broad feature is generally
present in absorption in radio galaxies and in emission in quasars,
and the tail of the feature can affect the continuum at 12~$\mu$m.  In
addition, due to the effects of dust extinction, for the same $L_{\rm
  bol}$ quasars can have up to twice as high values of $\nu L_{\nu 12
  \mu \rm m}$ than radio galaxies (e.g. \citealt{2009ApJ...706..184M,
  2008ApJ...685..160N}).  These two effects could moderately bias the
inferred values of $L_{\rm bol}$ by introducing a higher offset in
$\nu L_{\nu \rm 12\mu m}$ for quasars in relation to radio
galaxies. This implies that if the radio galaxies have high
extinctions, their luminosity at $12\,\mu \rm m$ is underestimated,
and therefore $L_{\rm bol}$ is in fact higher. This would in turn
bring the radio galaxies closer to the quasars, reducing the scatter
of the data points on the right panel of
Figure~\ref{fig:l151_l12}. \cite{2008ApJ...688..122H} quantify how
much extinction and the silicate feature can cause the luminosity of
quasars and radio galaxies to deviate from each other for wavelengths
up to $10\,\mu \rm m$. They show that at these wavelengths radio
galaxies can be on average 4 times less luminous than quasars due to
extinction. Therefore, if the radio loud galaxies in the sample suffer
high extinction ($\rm A_{V}\sim 50$), $L_{\rm bol}$ for the whole
bottom envelope would shift upwards, resulting in jet efficiency
values of $\eta\approx 0.3$ with $f=10$. Further work on this issue
will be presented in Fernandes et al. (in prep.)

One other matter of caution is the possibility of nonthermal
contamination of the mid-IR emission due to synchrotron emission from
the radio lobes or a core/jet component. Even though this is not a
major concern for our sample as nearly all sources are FRIIs and the
non-thermal contamination is in general a negligible fraction of the
emitted flux density at mid-IR frequencies, when Doppler-boosted or in
the case of steep-spectrum lobes lying within the Spitzer beam,
synchrotron emission from contaminant components can represent a small
fraction of the thermal emission (e.g. \citealt{2005ApJ...629...88S},
\citealt{2007ApJ...660..117C},
\citealt{2008ApJ...678..712D}). However, given the brightness and
radio spectral indices of the sources in our sample this is a minimal
effect.

It is an open question whether there is also a radio quiet envelope
and a real gap of objects beyond this region (to the left of region B
in the plot).  This is difficult to infer as the contribution from
stars begins to contaminate the radio emission at such low radio
luminosities. For instance, a typical powerful starburst with a star
formation rate of massive stars ($M\geq 5\rm \,M_{\sun}$) of $250\,\rm
M_{\sun}/yr$, equivalent to a total star formation rate of $~1000\,\rm
M_{\sun}/yr$ \citep{1992ARA&A..30..575C} assuming a Salpeter initial
mass function, is capable of producing radio luminosity densities of
the order of $\sim 5\times10^{23}\rm \,W\,Hz^{-1}\,sr^{-1}$ (vertical
dotted line in Figure~\ref{fig:l151_l12}). Moreover, the most powerful
starbursts observed, with a star formation rate of massive stars of
the order of $1000\,\rm M_{\sun}/yr$ equivalent to a total star
formation rate of $~4000\,\rm M_{\sun}/yr$, produce star-related radio
luminosity densities of $\sim 2\times10^{24}\rm
\,W\,Hz^{-1}\,sr^{-1}$, of the same order as luminosities produced by
weaker radio galaxies (vertical dot-dashed line in
Figure~\ref{fig:l151_l12}). Deep high-resolution multi-frequency radio
observations are required to cleanly distinguish AGN from purely
star-forming galaxies.

\section{Summary}

We have studied the relations between $L_{\nu \rm 151MHz}$, $\nu
L_{\nu \rm 12\mu m}$ and $L_{\rm [OII]}$ for a sample of 27
radio-selected galaxies (at $z\sim 1$), independent of evolutionary
effects due to redshift, and conclude that these are positively
correlated.

- A positive correlation between $L_{\nu \rm 151MHz}$ and $L_{\rm
  [OII]}$ confirms the previously known relation for larger samples,
supporting the idea that a link exists between the origin of the radio
jets and the source of the narrow lines.

- A positive correlation between $\nu L_{\nu \rm 12\mu m}$ and $L_{\rm
  [OII]}$ suggests that there is a common emission source that excites
the gas clouds in the NLR and the circumnuclear dust. This is
consistent with accretion onto the central black hole being
responsible for both forms of excitation.

- A positive correlation between $L_{\nu \rm 151MHz}$ and $\nu L_{\nu
  \rm 12\mu m}$ indicates that radio loud AGN have a high mid-IR
emission. Mid-IR emission in AGNs have a thermal component due to dust
that absorbs radiation from the accretion disc and re-radiates it at
these wavelengths. Assuming that the non-thermal contamination is not
relevant in our sample (see discussion in Section 4.4), this
correlation translates into a relationship between jet power and
accretion rate, which implies a common mechanism linking these two
physical properties (e.g. \citealt{1991Natur.349..138R}).

In addition, by adding a sample of bright optically selected quasars
we populated the $\nu L_{\nu \rm 12\mu m}$ vs $L_{\nu \rm 151MHz}$
plot and found that the objects span a diagonal region parallel to the
correlation found for the RGs. Thus, although the strength and slope
of the correlation seen in radio-selected samples is affected by
selection effects, there is a real gap of objects with bright $L_{\nu
  \rm 151MHz}$ and low $\nu L_{\nu \rm 12\mu m}$, which shows that for
a given jet power there is a minimum accretion rate. This implies that
there is a maximum efficiency with which accreted energy can be
converted into jet power, and that this efficiency is of order unity.

\section{Acknowledgements}

 We thank the anonymous referee for comments and suggestions that have
 greatly improved this paper. CACF is supported by the Foundation for
 Science and Technology (FCT-Portugal) through grant
 SFRH/BD/30486/2006. MJJ is supported by an RCUK fellowship. AMS is
 supported by an STFC post-doctoral fellowship. This work is based (in
 part) on observations made with the Spitzer Space Telescope, which is
 operated by the Jet Propulsion Laboratory, California Institute of
 Technology under a contract with NASA. This research has made use of
 the NASA/IPAC Extragalactic Database (NED) which is operated by the
 Jet Propulsion Laboratory, California Institute of Technology, under
 contract with the National Aeronautics and Space Administration.

\end{document}